\documentclass{aa}
\usepackage{graphics}

\def\ltapprox{\raise 2pt \hbox {$<$} \kern-1.1em \lower 5pt \hbox {$\approx$}}
\def\ltsim{\raise 2pt \hbox {$<$} \kern-1.1em \lower 4pt \hbox {$\sim$}}
\def\gtsim{\raise 2pt \hbox {$>$} \kern-1.1em \lower 4pt \hbox {$\sim$}}

\begin{document}

\title{Anisotropic inverse Compton scattering in powerful
radio galaxies: the case of 3C 295}

\author{G. Brunetti\inst{1,2}
\and M. Cappi\inst{3}
\and G. Setti\inst{1,2} 
\and L. Feretti\inst{2} 
\and D.E. Harris\inst{4} }

\offprints{G.Brunetti, c/o Istituto di Radioastronomia del CNR, via Gobetti
101, I--40129 Bologna, Italy; gbrunetti@astbo1.bo.cnr.it}

\institute{Dipartimento di Astronomia, 
via Ranzani 1, I--40126 Bologna, Italy
\and
Istituto di Radioastronomia del CNR,
via Gobetti 101, I--40129 Bologna, Italy 
\and
Istituto TeSRE-CNR, Via Gobetti 101, I--40129 Bologna, 
Italy
\and
Harvard-Smithsonian Center for Astrophysics, 
60 Garden st., Cambridge, MA 02138, USA}

\date{}

\abstract{
Inverse Compton (IC) scattering of nuclear photons
with relativistic electrons in the lobes of powerful radio 
galaxies and quasars 
can give detectable extended X--ray emission from
the radio lobes if relativistic
electrons with a Lorentz factor $\gamma < 300$ 
are present (Brunetti, Setti, Comastri 1997).
In general these electrons are not detected since they emit 
synchrotron radiation at frequencies below the radio band,
so that the study of this effect provides a unique tool 
to measure the energy distribution of the electron population 
in the radio lobes at $\gamma < 1000$ energies.
In this paper we reanalyze the {\it Chandra} observation
of the powerful and compact radio galaxy 3C 295 for which
the IC scattering of nuclear photons is expected to
be an important mechanism.
We find strong evidence for extended and asymmetrical
X--ray emission associated with the radio 
lobes in the energy band 0.1--2 keV.
We show that both the luminosity and 
morphology of the extended X--ray emission associated
with the radio lobes,  not compatible with
other X--ray mechanisms, can be best interpreted by the
IC scattering with nuclear photons.
We also show that the relativistic electron energy
distribution obtained from the synchrotron radio
emission can be extrapolated down to $\gamma \sim
100$ thus providing a first direct evidence on the
electron spectrum in the lobes down to lower energies.
}

\maketitle

\section{Introduction}

X-ray observations of samples of radio galaxies with
past generation satellites have considerably increased
our knowledge on their X--ray properties 
(e.g. Crawford \& Fabian 1995, 1996; Canosa et al. 1999;
Hardcastle \& Worrall 1999; Worrall 1999;  
Sambruna, Eracleous \& Mushotzky 1999; Capetti et al. 2000).
Basically, on large scales the X--rays are dominated 
by thermal X--ray emission from the clusters 
in which radio sources are generally embedded, 
while thermal emission from hot coronae of the host 
galaxies and non--thermal emission
from the nuclear/jet regions dominate
the X--rays on the smaller scales.
Due to the poor resolution of the X--ray observatories, 
the diffuse emission from the cluster hot gas 
made it difficult to study the extended X--rays
from the lobes and hot spots
of radio galaxies and quasars.
As a consequence, only in a few cases  
IC fluxes from the radio hot spots 
(Harris, Carilli \& Perley 1994) and 
lobes (Feigelson et al.1995; Kaneda et al.1996;
Tsakiris et al.1996; Tashiro et al.1998; 
Brunetti et al.1999; Tashiro et al.2000) 
were detected.

The advent of {\it Chandra}
will probably cause significant progress in the 
knowledge of radio galaxies and quasars as
demonstrated by the promising
first results on X--rays from jets and hot spots  
(Chartas et al. 2000;
Harris et al. 2000; Schwarts et al. 2000; 
Wilson Young \& Shopbell 2000,2001).

Due to the combination of high spatial resolution and sensitivity
{\it Chandra} can also study in detail kpc scale
diffuse X--ray emission originating from
the lobes of radio galaxies and quasars
where electrons can radiate via IC 
scattering of CMB photons (e.g. Harris \& Grindlay 1979), 
and/or nuclear photons (Brunetti, Setti \& Comastri 1997).
\begin{figure*} 
\resizebox{\hsize}{!}{
\includegraphics{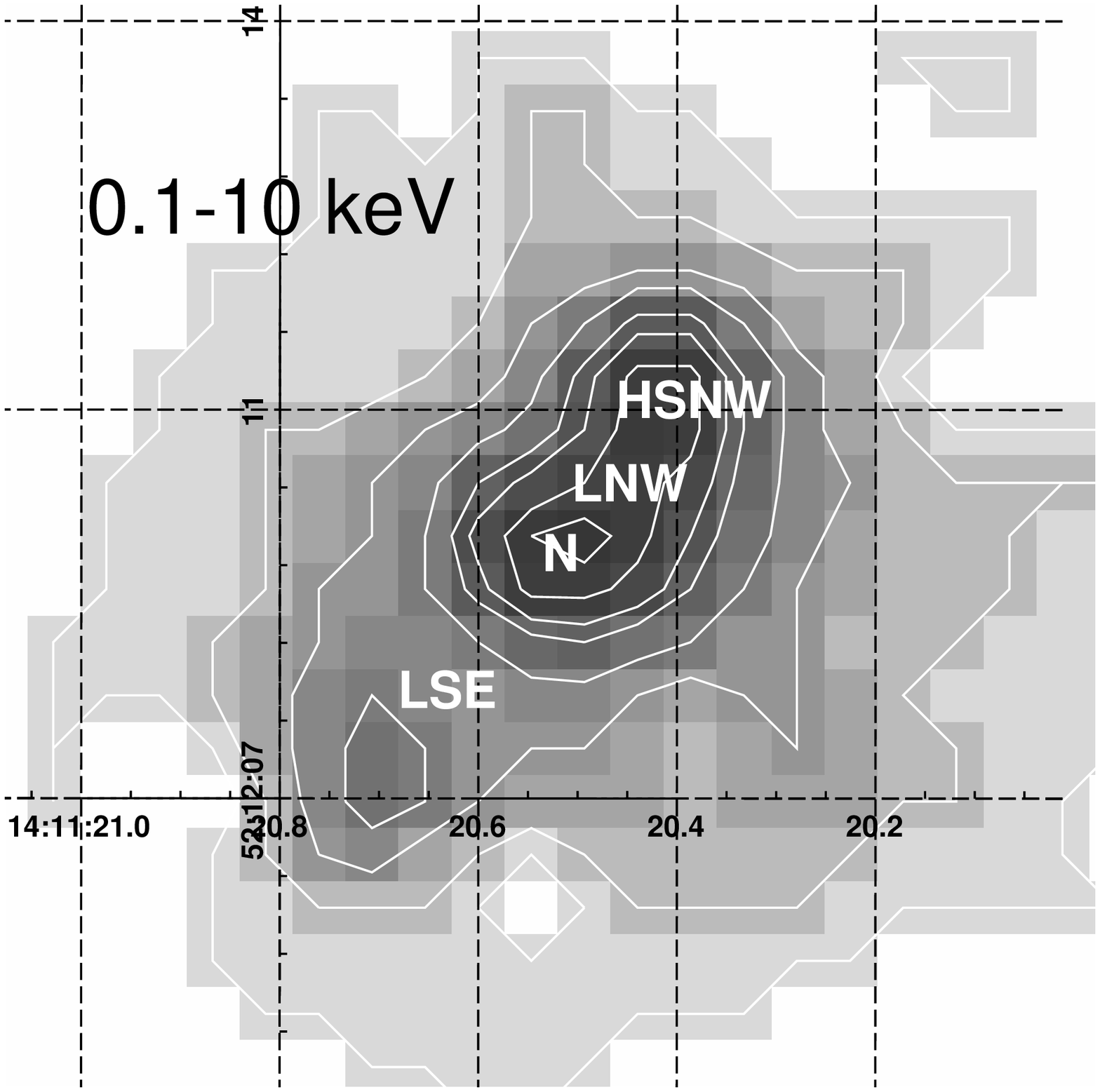}
\includegraphics{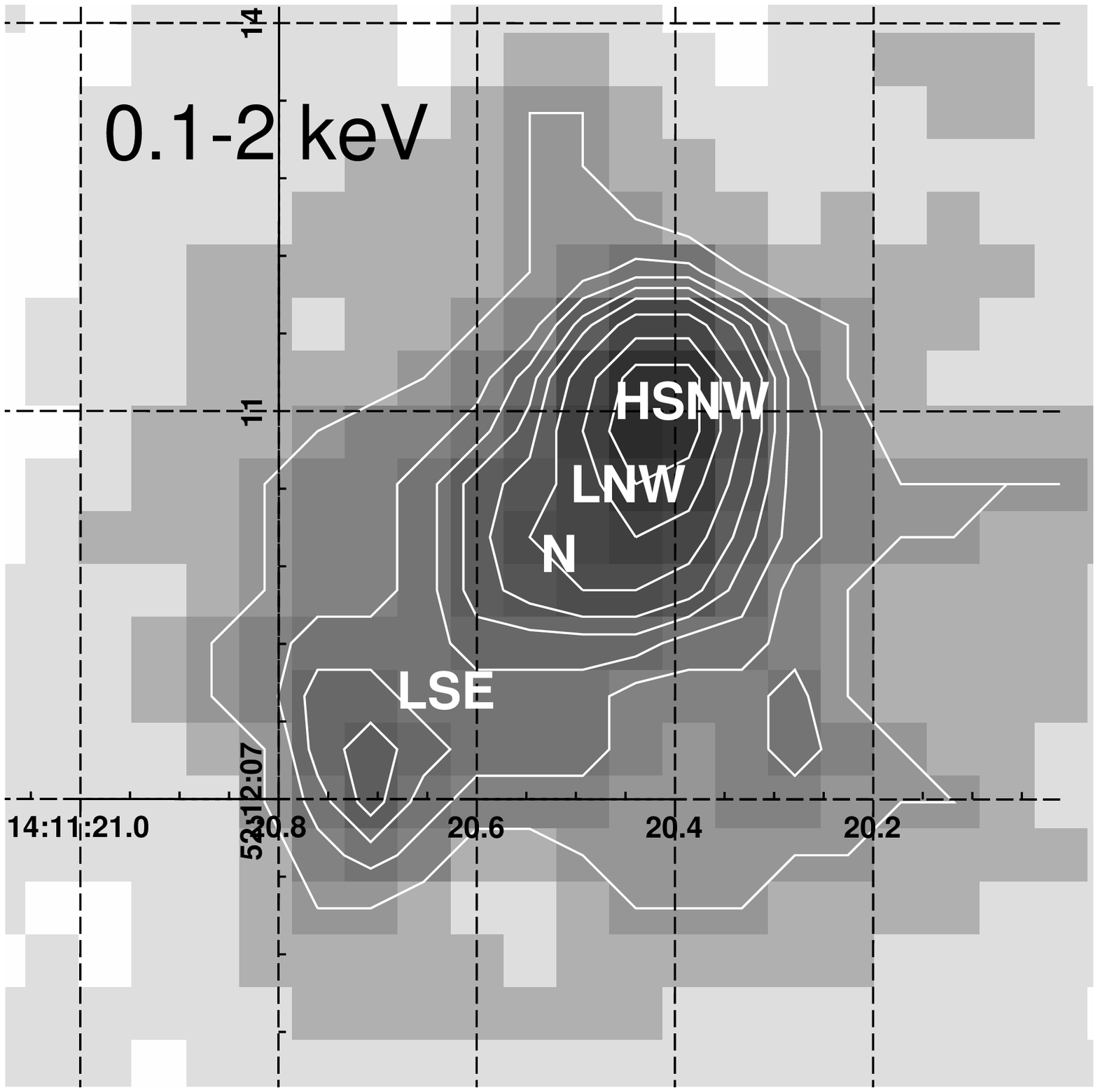}
\includegraphics{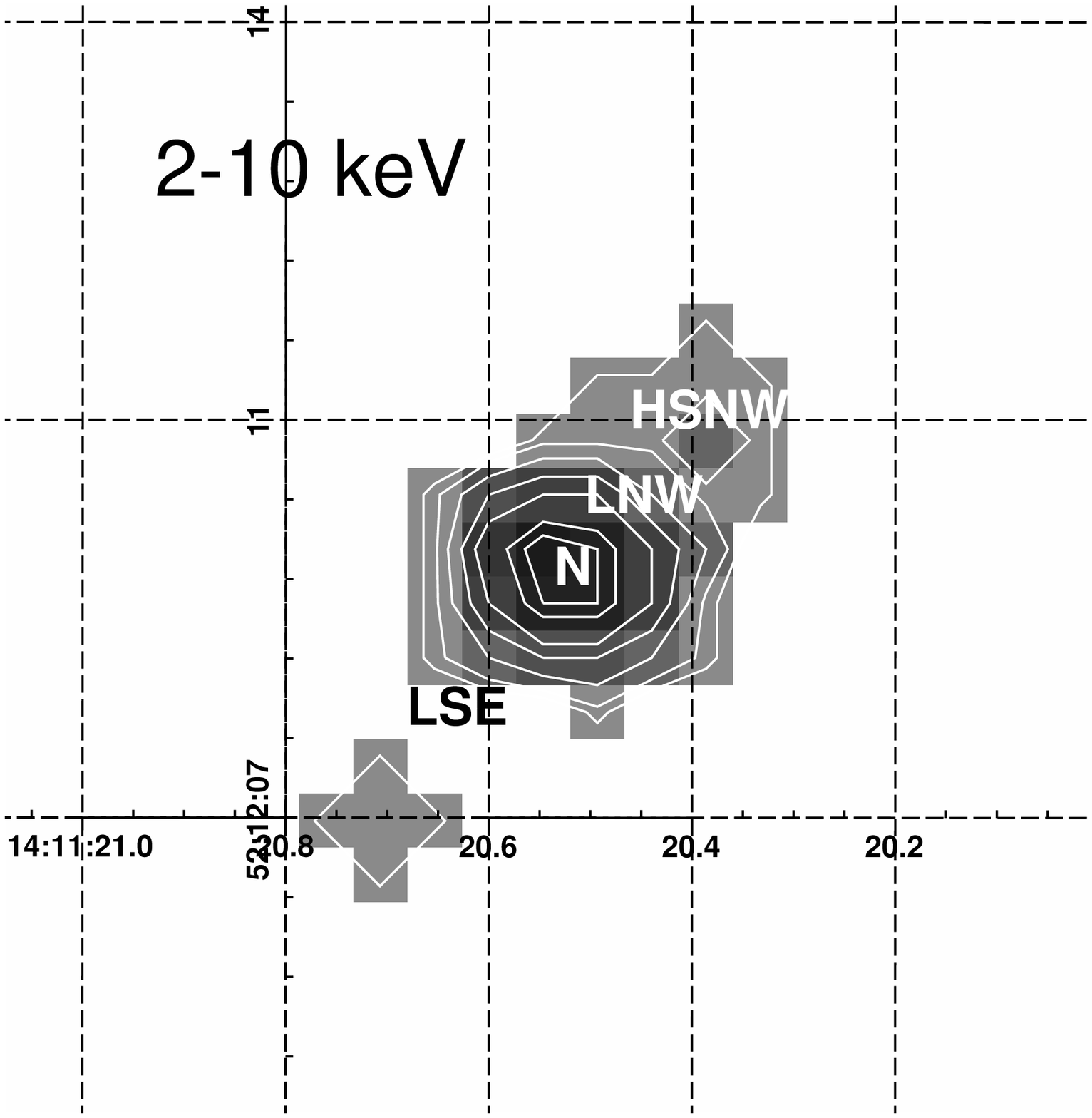}}
\caption{
{\bf Left panel}: 0.1--10 keV image of 3c 295
smoothed with a Gaussian of $\sigma = 0.5$ arcsec.
The reported contours are 3, 5, 7, 10,
12, 15, 17 and 17.4 cts/px (1 px=$0.5\times 0.5$ arcsec).
The text labels indicate the 
different X--ray components: nuclear emission
(N), northern hot spot emission (HSNW), and
lobe emission (LSE, LNW).
{\bf Central panel}: 0.1--2 keV image of 3C 295
smoothed with a Gaussian of $\sigma = 0.5$ arcsec.
The reported contours are 4.5, 5.5, 6.5, 7.5, 
9, 11, 13, 15 cts/px.
{\bf Right panel}: 2--10 keV image of 3C 295
smoothed with a Gaussian of $\sigma = 0.5$ arcsec.
The reported contours are 1.2, 2, 3, 4, 6, 7, 8
cts/px.}
\end{figure*}
While the IC emission in the case of CMB seed photons is 
contributed by typically the less energetic part of the 
radio emitting electrons 
(Lorentz factor $\gamma \sim 10^3$), 
the X--rays from IC scattering of 
the nuclear far--IR/optical photons is mainly powered
by $\gamma=100-300$ electrons 
whose synchrotron emission typically falls
in the undetected hundred kHz frequency range.

In the commonly accepted scenario, 
the relativistic plasma filling the lobes of 
FRIIs is believed to be released by the radio
hot spots in which it is probably re--energized by
shocks (e.g. R\"{o}ser \& Meisenheimer 1986). 
In general, the main requirement for the initial energy of the 
electrons in order to achieve an efficient 
reacceleration is $\gamma >> 1$ (e.g. Bell 1978), but it 
depends on the assumed acceleration scenario 
(see Eilek \& Hughes 1990 and ref. therein); 
this makes the measure 
of the relativistic electron spectrum at lower energies
particularly useful to
constrain the acceleration mechanisms in radio sources. 
Due to the surrounding emission from the radio lobes and
synchrotron self absorption, 
low frequency radio observations of the 
hot spots are difficult.
However, there are some evidences for synchrotron emission 
by $\gamma \geq 400$ electrons (Meisenheimer 1989; 
Carilli et al. 1991,1999).
In addition, current models consider the hot spots to be over 
pressured with 
respect to the
surrounding cocoon (e.g. Keiser \& Alexander 1997),  
so that the relativistic plasma would adiabaticlly
expand when it leaves the hot spot region
(e.g. Carilli 1991).
As a consequence, any low energy cut--off in the hot spots' 
electron population would be further shifted to lower energies
so that, in principle, the presence of $\gamma \geq 10-100$ 
electrons in the radio lobes cannot be ruled out.

In order to test the presence of these particles 
and to measure their energy distribution,  
observations of X--ray 
fluxes from IC scattering of
nuclear photons are particularly useful.
Due to the dilution of the nuclear
photon flux with distance, the X--ray emission is expected
to be more concentrated toward the innermost parts
of the radio lobes at variance with the expected distribution
of the X--rays from the standard IC 
scattering of CMB photons.
Furthermore, due to the geometrical 
configuration, the IC scattering is anisotropic so that,  
if the radio axis of a symmetric double lobed radio galaxy 
is inclined with respect to the plane of the 
sky, the far lobe should appear more luminous than 
the near one (Brunetti et al. 1997).
The X--ray spectra may provide important information; indeed, 
the presence of a low energy flattening of the electron
distribution (or a cut--off) at energies $\gamma \leq 100$ 
results in a somewhat flatter spectrum than that 
extrapolated from higher energies (radio spectrum and/or
from IC scattering of the CMB photons).
Furthermore, in the case of a symmetric double lobed
radio galaxy, the soft X--ray spectrum of the far lobe 
should appear slightly harder that that of the near one 
(Brunetti 2000).

Possible evidence for this emission has been found in the case 
of the powerful radio galaxy 3C 219 by a
relatively deep ROSAT HRI observation corroborated 
by a combined ROSAT PSPC and ASCA spectral analysis
(Brunetti et al. 1999)\footnote{During the preparation of the paper
a possible evidence for IC scattering of nuclear photons
has also been suggested by the {\it Chandra} observation of 
the high--z (z=1.78) radio galaxy 3C 294 (Fabian et al., 2001)}.

In this paper we reanalyze the {\it Chandra} observation
of the powerful radio galaxy 3C 295 giving evidence for 
diffuse X--ray emission which supports the
IC scattering of the photons from the hidden quasar.

\noindent
$H_0 =50$ km s$^{-1}$ Mpc$^{-1}$ and $q_0 =0.5$ are
assumed throughout; at the distance of the radio galaxy
1 arcsec corresponds to 6.9 kpc.

\section{Target and Data analysis}

The powerful radio source 3C 295 is identified with a giant 
elliptical (cD)
galaxy at the center of a rich cluster (z=0.461).
The radio image (Taylor \& Perley 1991) shows a classical double 
lobed morphology with hot spots and a very high 
total radio power of $\sim 10^{36}$ erg $s^{-1}$ Hz$^{-1}$ 
at 178 MHz.
The X-ray data obtained with previous instruments 
({\it Einstein} Observatory: Henry \& Henriksen 
1986; ASCA: Mushotzky \& Scharf 1997;  ROSAT: 
Neumann 1999) permitted the study of only 
the cluster emission.

The calibration observation of 3C 295 was performed
on 1999, August 30th for an elapsed time of $\sim$20 ks.
The target was near the aim point on the S3 ACIS chip.
The data were cleaned and analyzed using the
{\it Chandra} Interactive Analysis of Observations
(CIAO) software (release V1.1, Elvis et al. 2000, in
preparation, see also http://asc.harvard.edu/cda/).
The data were first filtered to include only the
standard event grades 0,2,3,4 and 6, and energies
between 0.1--10 keV.
All hot pixels and bad columns were removed.
Time intervals with large background rate
(i.e. larger than $\sim 3\sigma$ over the quiescent state)
were removed, yielding a screened exposure time
of 18364 s.
Images obtained between 0.1--10 keV, 0.1--2 keV and 2--10 keV
are shown in Fig.1.
These clearly show that the X--ray emission from 3C 295
consists of several components:
a nuclear point source which shows up at energies $>$ 2 keV, 
diffuse cluster emission, diffuse lobe emission and
clear hot spot emission (at least in the north--west lobe).

Details on the cluster and hot spot emission are given in 
Allen et al.(2000) and Harris et al.(2000), respectively. 
Hereinafter, reported errors correspond to intervals at
90$\%$ confidence level for one interesting
parameter ($\Delta \chi^2 = 2.71$) unless specified otherwise.

\subsection{The nuclear source}

Harris et al.(2000) fitted the strong nuclear source with
a power law model in the range 0.5--7 keV obtaining
an inverted energy power law with a 
photon index $\Gamma = 0.2 \pm 0.3$ and a luminosity
of $\sim 10^{44}$ erg s$^{-1}$ between 0.2--10 keV.

Stimulated by these results, 
since the nucleus of a radio galaxy 
is expected to be highly absorbed 
(Barthel 1989; Ueno et al. 1994),  we attempted 
a fit of the nuclear spectrum by including an absorbed
power law model.
We did not restrict ourselves to 0.5--7 keV 
(if so confirming Harris et al.2000 results)
but used all the data from 0.1--10 keV requiring 
$\geq 10$ counts/bin and $S/N >$3.
Fits have been performed by extracting circular
regions centered on the source peak
with 1 arcsec radius (see Fig.1) and 
with the 'background' spectrum extracted
from circular region close to the source
accounting for the cluster emission.
Fits with different extraction radii (e.g. in the
range 0.5--1 arcsec) and different backgrounds 
gave similar results.
The results are reported in Fig.2 while the main
parameters from the fit are reported in Table 1.

\begin{figure}
\includegraphics{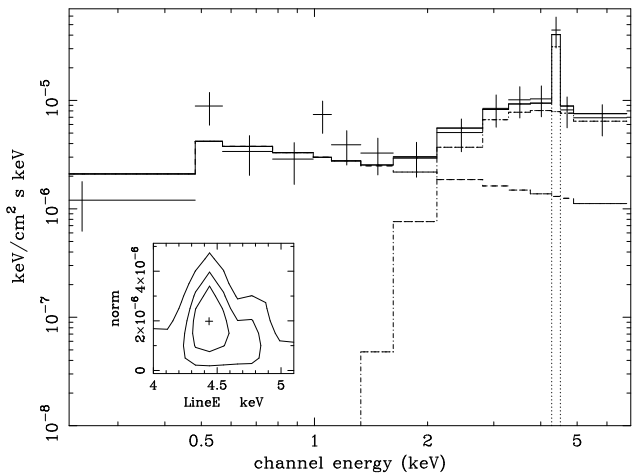}
\vspace{7 cm}
\caption{
The photons extracted from a circular region of
1.5 arcsec radius centered on the nuclear source
are fitted with a power law model (dashed line), plus
an absorbed power law (dot--dashed line) and an
iron line component (dotted line).
In the small panel 
68, 90 and 99 \% confidence levels of the
iron line normalization vs energy center 
are reported.
The parameters of the fit are given in Table 1
(third row).}
\end{figure}

\begin{table*}
\caption{Results from spectral fits (f indicates a frozen 
parameter)}
\begin{tabular}{lllllllll}
\hline 
\noalign{\smallskip}
N$_{\rm H}^{\rm a}$ & $\Gamma^{\rm b}$ & $kT^{\rm c}$ & N$_{\rm H}^{\rm d}$ 
& $\Gamma^{\rm e}$&  
$E_{\rm line}^{\rm g}$ & $\sigma_{\rm line}^{\rm h}$ & $\chi^{2}/dof^{\rm i}$ 
\\
\noalign{\smallskip}
\hline
\noalign{\smallskip}
1.34 (f) 
& 0.6$\pm$0.3 & 
-- & 0.0 (f) 
&--&
-- 
& -- & 24.0/13 \\
1.34 (f) & 
 1.6 (f) 
& --  & 11.9$\pm$7.3 & 3.6$\pm$2.0 &
4.5 (f) & 0.0 (f) &
9.7/10 \\
1.34 (f) 
& 1.6 (f)
& -- & 6.2$\pm3.2$ & 1.9 (f) & 4.5 (f) & 0.0 (f) &
10.5/11 \\
1.34 (f) 
& -- & 5.0 (f) & 10.3$\pm$4.8
& 3.1$\pm$0.6& 4.5 (f)  & 0.0 (f) & 8.5/10 \\
1.34 (f)  
& -- &
5.0 (f) & 6.6$\pm$2.1 & 
2.0 (f) & 4.5 (f) & 0.0 (f) & 10.4/11 \\
\noalign{\smallskip}
\hline
\noalign{\smallskip}
\end{tabular}

$^{\rm a}$ Equivalent hydrogen column density (units of $10^{20}$ cm$^{-2}$) 
fixed at Galactic value\\
$^{\rm b}$ Photon spectral index of the scattered power law\\
$^{\rm c}$ Temperature of the Raymond--Smith model (keV)\\
$^{\rm d}$ Column density of the absorber ($10^{22}$ cm$^{-2}$) \\
$^{\rm e}$ Photon spectral index of the absorbed power law\\
$^{\rm g}$ Iron line energy (keV) \\
$^{\rm h}$ Iron line width (eV) \\
$^{\rm i}$ Total $\chi^2$ and degrees of freedom 
\end{table*}

The spectrum is well fitted 
by an absorbed power law model
(which accounts for the hard emission)  
plus either a Raymond--Smith model ($kT \sim 5$ keV)
or an unabsorbed power law model ($\Gamma \sim 1.6$)
which accounts for the soft emission.
The poor statistics do not allow us to discriminate
between the Raymond--Smith and the second 
power law from the spectral 
analysis alone.
The soft X--ray component could be due to
an enhancement of the cluster thermal emission
in the extracted region, to nuclear
scattering from cold electrons in the vicinity of the
AGN, or to the non--thermal IC scattering from the 
innermost lobes.

By leaving the photon index of the absorbed power law
free to vary we obtain best values of  
$\Gamma \simeq 3.1 \pm 0.6$ 
and $N_{\rm H} \simeq 1\pm 0.4 \times 10^{23}$ cm$^{-2}$
(i.e. $\sim 10^3$ times the Galactic value)
which correspond to a 0.1--10 keV deabsorbed 
luminosity of $\sim 10^{46}$ erg s$^{-1}$.
Nevertheless, the very steep photon index 
obtained in this fit might be due to the poor statistics
at higher energies, so that we also attempted fits 
with an absorbed power law model of 
frozen photon index $\Gamma$ =1.9 and 2.0 obtaining
equally good fits; in this case the fitted 
column density is $N_{\rm H} \simeq 6-8 \times 10^{22}$ cm$^{-2}$
and the 0.1--10 keV luminosity reduces to $\sim 6 \times
10^{44}$ erg s$^{-1}$.  

The presence of a powerful hidden AGN absorbed by a 
high column density is further corroborated by 
the marginal detection 
of an iron line in the nuclear spectrum (Fig.2) 
centered in the range 4.3--4.6 keV 
(rest frame 6.3--6.7 keV) with 
EW = $660^{+1400}_{-180}$eV.

In order to avoid possible statistical biases due to the 
adopted binning procedure 
we have also analyzed the nuclear spectrum with
an unbinned fitting procedure.
A statistical test that does not require binning
entails minimizing the 'C-statistics' 
(Cash 1979; see also Weaver 1993).
This statistics cannot
be applied to background--subtracted data, 
thus to fit the nuclear data we use a model 
consisting of the source plus 
the 'background' as obtained 
in the previous fits.
The source is modeled with  
a power law (with a frozen photon index 
$\Gamma=1.6$ absorbed at the Galactic value) 
plus an absorbed power law
(with $\Gamma$ and $N_{\rm H}$ free to vary).
The unbinned fitting procedure gives results fully 
consistent with those from the binned procedure
confirming the presence of absorption excess 
($N_{\rm H} > 6 \cdot 10^{22}$cm$^{-2}$).

\subsection{The diffuse X--ray emission from the 
radio lobes}

Both the 0.1--10 keV and 2--10 keV images (Fig.1) 
are dominated by the strong central source, this combined with
the presence of the luminous northern X--ray hot spot 
at $\sim 2$ arcsec distance 
makes problematic the analysis of possible diffuse
emission associated with the radio lobes.

As discussed above the nuclear source is
absorbed at low energies and does not significantly
contribute to the X--ray emission in the 0.1--2 keV band.
The 0.1--2 keV image of the central part of the
cluster is shown in Fig.1. Diffuse X--ray emission  
with a double lobed structure exceeding the 
cluster is clearly detected.
Although the brightness of the 
X--ray lobes is from 1.5 to 5 times larger than 
that of the surrounding cluster, 
a subtraction of the cluster emission 
leads to a better quantitative estimate.
We first performed a number of fits of the cluster 
brightness profile
using a $\beta$--model and excluding a circular region of radius
2--3 arcsec centered on the nucleus of 3C 295.
Both the 0.1--2 keV and 0.1--10 keV unsmoothed
images have been used in the
fitting procedure. 
The X--ray distribution is fitted in different azimuthal
angle range: 
in two quadrants of 90$^o$ opening angle in the 
direction perpendicular to the radio axis, in the
direction of the radio axis and 
over 0--360$^o$.
We have subtracted the 
best fitted $\beta$--model (with core radius
4.0 arcsec and $\beta = 0.58$) from the 0.1--2 keV 
total unsmoothed image.
The result is shown in 
Fig.3 where the X--ray brightness distribution is 
superimposed on the 8 GHz VLA radio 
image with a resolution of 0.25 arcsec
(kindly provided by Taylor \& Perley).
Different $\beta$--models (with core radius between
4.0--4.4 arcsec and $\beta =0.54-0.58$, consistent with
the range of parameters from Neumann 1999 and 
Harris et al.2000) gave similar results to within $\sim 10 \%$.

The resulting morphology of the diffuse X--ray emission (Fig.3) is 
double lobed with the X--rays coincident with the radio 
lobes, thus a non--thermal
mechanism is by far the most likely to apply.
Furthermore, both recent and past X--ray 
observations of radio sources in clusters 
have not revealed thermal X--ray
excesses in coincidence with the radio
lobes and jets, but rather deficits (Carilli, Perley, Harris 1994;
Fabian et al. 2000; McNamara et al. 2000).
Moreover, at variance with our findings, 
heating of the intergalactic medium 
by the cocoon of the radio galaxies would basically produce  
edge--brightened X--ray emission (e.g. Kaiser \& Alexander 1999).
Based on these considerations in the following section
we investigate only non--thermal mechanisms as the source
of the diffuse X--rays from the radio lobes.

The northern lobe in Fig.3 
appears to be slightly more extended 
than the radio brightness distribution.
In some regions, this is not only caused by the larger
{\it Chandra} PSF but it could be a real effect tracing the 
presence of lower brightness radio extension not visible 
at 8 GHz, but detected in 1.4 GHz radio images  
(see Harris et al.2000 Fig.1, and ref. therein).
The X--ray emission associated with the 
northern lobe in Fig.3 is extended 
also in the direction perpendicular to the radio axis.
In the distance range 0--1.5 arcsec from the nucleus
we measure $48\pm11$ net counts 
between 0.5--1 arcsec from the radio
axis, whereas a narrow beam  
along the radio axis would give $18\pm6$ net counts
(if normalized to the observed counts 
between 0--0.5 arcsec).
From this analysis we find that the intrinsic width of the
northern X--ray lobe is $>0.8$ arcsec 
when averaged over a radial distance of 0 to 1.5 arcsec.

We notice that the resulting X--ray emission is very asymmetrical
with the northern lobe being the most luminous.
\begin{figure}
\resizebox{\hsize}{!}{
\includegraphics{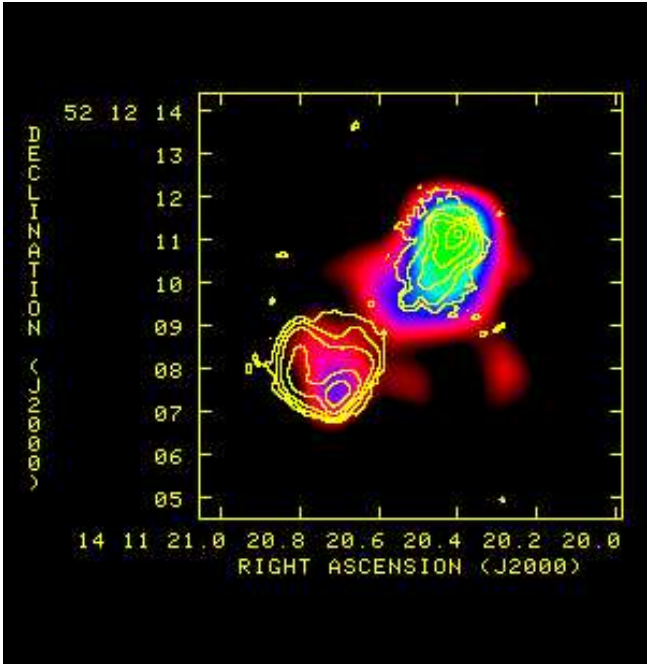}
}
\caption[]{
The 0.1--2 keV X--ray image (color scale), obtained
after the subtraction of the cluster emission, is shown
superimposed on the 8 GHz VLA radio map (contours).
The X--rays have been shifted by 0.3 arcsec
to match the position of the nucleus, measured on the 
0.1--10 keV image, with that of the radio core.
The radio contours are: 0.2, 1, 5, 25, 125, 500 mJy/beam
(beam = 0.25$\times$0.25 arcsec).}
\end{figure}
In order to better quantify the asymmetry
of the diffuse X--ray flux associated
with the radio lobes we have subtracted the luminous
northern hot spot from the unsmoothed 0.1--2 keV image
by adopting either an unresolved source model (PSF), 
or slightly resolved models resulting
from the convolution of optically thin
spherical hot spots 
with 0.2-0.4 arcsec diameter with the PSF; the result is 
reported in Fig.4.
We measure $\sim 150-240$ net cts between 0.1--2 keV
in excess of the residual 'background' and 
coincident with the radio lobes.
This depends on the details of the northern hot spot subtraction,
240 cts corresponding to no subtraction, 150 cts to 
the subtraction of a 0.4 arcsec hot spot (i.e., 
the hot spot would account for up to 90 net counts in the 
0.1--2 keV band).
Furthermore, we find that the ratio between the 0.1--2 keV net cts 
associated with the two radio lobes is 
$2.2\pm0.4$ and $3.9\pm0.8$ in the case of maximum hot spot subtraction 
and no hot spot subtraction, respectively.
If a fraction of the X--rays coincident
with the southern lobe is contributed by the SSC process 
from the hot spot (Harris et al.2000)
the ratio between the X--ray fluxes from
the northern and the southern lobes would be 
further increased.
After subtraction 
of the northern hot spot and of 
the total X--ray flux coincident with
the position of the southern radio hot spot such a ratio
becomes $3.4\pm0.9$; in the following we consider
this as the upper value of the X--ray asymmetry.

To obtain spectral constraints of the diffuse X--rays 
from the lobes, we considered a region of
11 arcsec$^2$ coincident with  the radio lobes 
(i.e. 44 px, 1 px=0.5$\times$0.5 arcsec).
We fitted the spectrum  
with a simple power law model 
(Galactic absorption)
by excluding a circular region of 0.5 arcsec radius
centered on the northern hot spot; the 
'background' spectrum was extracted 
from cluster regions close to the radio source.
The resulting photon index is $\Gamma = 1.44^{+0.20}_{-0.21}$; 
we also stress that the fitted
spectrum would steepen if $N_{\rm H}$ is allowed to be larger
than the Galactic value.
The corresponding 0.1--10 keV luminosity is 
$\sim 5-7 \cdot 10^{43}$erg s$^{-1}$ 
(the lower corresponding to $\Gamma = 1.6$).

\begin{figure}
\resizebox{\hsize}{!}{
\includegraphics{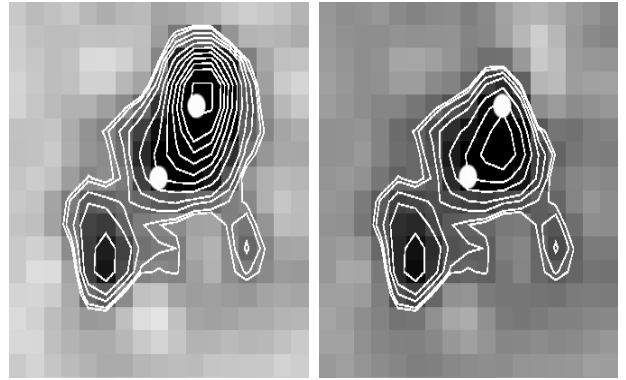}}
\caption[]{X--ray isophotes
superimposed on the 0.1--2 keV
{\it Chandra} pixel
map   
before the northern hot spot subtraction
(left panel) and after the subtraction
(right panel).
The subtracted hot spot model is a 
sphere of 0.4 arcsec diameter convolved
with the {\it Chandra} PSF and normalized
to 90$\%$ of the peak.
The images are smoothed with a gaussian of
$\sigma$ = 1 pixel, the position of the nucleus 
and of the X--ray northern hot spot are also shown.
The contours are 1.8, 2.0, 2.5, 3.0, 4, 5, 6, 7, 8, 9,
11.2 cts/px (1 px=$0.5 \times 0.5$ arcsec).}
\end{figure}

The radial profile of the diffuse X--rays coincident 
with the radio lobes is difficult to determine 
due to the small extension
of the lobes ($\sim 2$ arcsec) and to the presence of
the northern hot spot.
We have measured the counts on the
0.1-2 keV image in two regions of the northern lobe
of 0.75 arcsec$^2$ (3 pixels) at a radial distance 
from the nucleus of 0--0.5 arcsec and  
1--1.5 arcsec respectively.
We obtain a brightness ratio between the
1--1.5 and 0--0.5 region in the range 
0.7--1.6 depending on the hot spot subtraction
(1.6 corresponding to no subtraction).
We conclude that the diffuse X--rays in excess of
the northern hot spot emission have a
rather flat brightness distribution between
0--1.5 arcsec from the nucleus; this may 
further help us in the explaining of the origin
of the emission (Sect. 3.2).

\section{The IC scenario}

\subsection{Photon energy densities}

The spectral study of the nuclear emission (Sect. 2.1) 
strongly supports the presence of 
a highly absorbed powerful 
quasar in the nucleus of 3C 295 with a luminosity of
$\sim 10^{45}$erg s$^{-1}$ in the 
0.1--10 keV band.

By assuming quasar spectral energy distributions (SEDs) 
one has that the far--IR to optical luminosity
is typically $\sim 5-15$ times larger than the
soft X--ray luminosity with the flat spectrum
radio loud quasars having
the lower far--IR/X--ray ratio
(Sanders et al.1989; Elvis 1994).
While recent studies have shown that the far--IR photons in 
quasars are almost certainly due
to dust emission (Haas, Chini and Kreysa 1998; van Bemmel, 
Barthel and Yun 1999), combined radio and X--ray studies have
revealed the existence of a beamed plus an isotropic
(or quasi isotropic) X--ray component
in radio loud quasars (Brinkmann, Yuan and Siebert (1997) 
and ref. therein).
In the case of 3C 295 the X--ray beamed component is   
suppressed by transverse Doppler boosting and 
we are essentially 
measuring the (quasi)isotropic X--ray emission.
Therefore we adopt a ratio $\sim 10-15$
(more typical of steep spectrum radio loud
and radio quiet quasars)
and estimate a 
far--IR/optical luminosity 
of the hidden quasar $\sim 10^{46}$erg s$^{-1}$.

Since 3C 295 is a relatively 
compact radio source with a 
total extension of $\sim 30$ kpc, the
energy density due to the nuclear photons
is expected to be important over all the
radio volume; it is:

\begin{equation}
\omega_{\rm Q}(r) = { L_{\rm Q} \over { 4 \pi c r^2 }}
\simeq
3 \cdot 10^{-9}
{{ L_{\rm Q,46} }\over{ r_{\rm kpc}^2 }}
\,\,\,\,\,\,\,\,\,\,\,\,\,\,\,\,\,\,\,\,\,(cgs)
\label{qso}
\end{equation}

\noindent
$L_{\rm Q,46}$ being the far--IR/optical  
luminosity in units of $10^{46}$erg s$^{-1}$,
$r_{\rm kpc}$ the distance from the nucleus in kpc.

The high brightness of the radio lobes 
suggests that synchrotron--self--Compton (SSC) 
could be an important mechanism to produce
X--rays.
Indeed, Harris et al. (2000) suggested that a large
fraction, if not all, of the X--ray source 
coincident with the northern hot spot 
of 3C 295 is powered by such a mechanism.
\begin{figure}
\resizebox{\hsize}{!}{
\includegraphics{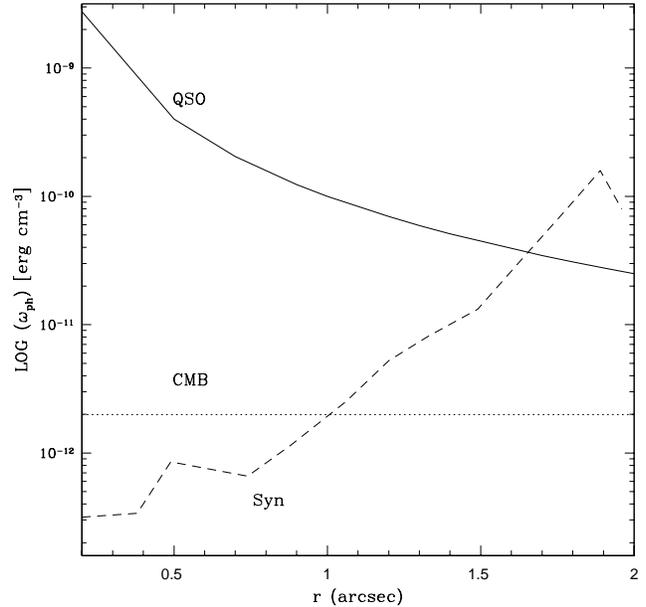}}
\caption[]
{The energy density due to the
nuclear photons (solid line) in the
northern lobe of 3C 295 is compared
with that due to synchrotron (dashed line)
and CMB photons (dotted line).
The luminosity of the hidden quasar is 
$L_{\rm Q}=10^{46}$erg s$^{-1}$ and
the synchrotron energy density
is calculated along the radio axis.}
\end{figure}
By assuming an axial symmetry of the 
emitting regions, we have obtained 
the synchrotron energy density in 
the northern radio lobe
along the direction of the radio axis
using the VLA radio image of Fig.3. 
The results are reported in Fig.5 together with
the quasar photon energy density (Eq.\ref{qso})
and with the CMB energy density.
We find that 
the energy density of the nuclear photons   
dominates the synchrotron and CMB 
up to the hot spot region ($\sim 1.6$ arcsec, 
or $\sim 11$ kpc from the nucleus), while the
synchrotron energy density dominates the CMB one for 
$r \geq 1$ arcsec (i.e. $\sim 7$ kpc).
As a consequence, the extended X--ray
emission detected between 5--10 kpc from the
nucleus is likely to be powered by 
the IC scattering of 
far--IR/optical quasar photons.
This conclusion is further reinforced by the 
considerations developed in the following
sub--section.

\subsection{X--ray brightness asymmetry and geometry}

As already stated in Sect. 2.2 the extended X--ray emission
associated with the radio lobes is asymmetrical, 
the integrated X--ray flux from the northern lobe being 
a factor $\sim 2-4$ larger than that from the southern one.
At radio wavelengths (Fig.3) the northern lobe appears
more compact and closer to the nucleus than the southern one, 
while the radio luminosity of the two lobes is similar.
The comparison between radio and X--ray morphologies 
can be used to rule out non--thermal mechanisms different 
from the IC scattering of nuclear photons as the main source
of the diffuse X--rays.

Indeed, by assuming the same constant ratio between magnetic
and electron energy densities in the two lobes,  
we calculate that the X--ray flux produced by IC scattering
of CMB photons in the southern lobe is expected to
be slightly larger than that in the northern lobe, 
at odds with the {\it Chandra} results. 
If nuclear photons were not involved, yet IC scattering was
the cause of X--rays, then the lobe 
asymmetry could only be reproduced by 
the {\it ad hoc} assumption that the ratio
between electron and magnetic energy densities
in the northern lobe is from $\sim$20
to $\sim$80 times 
larger than that in the southern one
(for an X--ray ratio 2 to 4 respectively).

It is interesting to notice that the SSC mechanism,
although unimportant for large scale emission
as previously shown, is likewise 
ruled out by the observed distribution of the
X--ray brightness $\Sigma_{\rm x}$.
From the standard SSC and 
synchrotron formulae (Gould 1979) one 
roughly finds
that, by assuming 
a constant ratio between relativistic electrons and
magnetic field energy densities, 

\begin{equation}
\Sigma_{\rm x} \propto  
\Sigma_{\rm r}^{ {{\delta+9}\over{\delta+5}} } l_{}^{ {{\delta+1}\over{
\delta+5}} }
\end{equation}

where $\Sigma_{\rm r}$ is the radio brightness (Fig.3), 
$l$ the thickness of the radio emitting volume
along the line of sight and $\delta$ the relativistic 
electron spectral index.
By assuming axial symmetry and $\delta \sim 3$
it follows that $\Sigma_{\rm x}$ should increase
by a factor $\sim 30$ between 0--0.5 and 1--1.5 arcsec
distance from the nucleus, at odds with the rather
flat observed $\Sigma_{\rm x}$ distribution.
To be consistent with the
data one would have to assume 
that the ratio of the energy densities in 
relativistic electrons to that in the
magnetic field varies by a factor \gtsim$10^3$
going from the outer to the inner regions mentioned
above.

On the contrary, as already anticipated in the
Introduction, the X--ray lobe asymmetry is a
natural consequence of the IC scattering
of nuclear photons.
As it is well known, in the IC process a 
relativistic electron scatters the incoming photons
in a narrow cone about its instantaneous
velocity vector and the highest energies of the
scattered photons are obtained in the head--on scatterings.
Since in the radio lobes the momenta of the relativistic
electrons are assumed to be isotropically distributed,
it follows that at any given energy of the scattered
photons there will be many more scattering events when the
electrons move toward the nucleus, 
i.e. the direction of the incoming photons, than 
when they are moving away:
the resulting IC emission will be enhanced towards the
nucleus and essentially absent in the opposite direction.
As a consequence, if the radio axis does not lie on the plane of the
sky, the smaller the angle between the axis and the line of sight,
the greater will be the difference in IC emission from two
identical lobes (see Figs. A2--A4).

\section{The Model}

We have applied the anisotropic IC scattering
equations reported in the Appendix to 3C 295.  
The main ingredients of the model are as follows:

{\it a) Geometry of the radio lobes}.
We have assumed that the radio volume
is symmetrical about the line joining the mid--points
of the weakest radio isophote (Fig.3)
at each fixed distance from
the nucleus and that the ratio between the energy 
densities in relativistic electrons and magnetic fields
is constant.

{\it b) Spectrum of the relativistic electrons}. 
As a starting point for the model it can be constrained
by the radio observations.
At low radio frequencies the spectrum of 3C 295 is dominated by the
emission from the radio lobes, while 
the emission from the hot spots contaminates the
lobe spectrum at higher frequencies, 
contributing $\sim 30 \%$ of the total flux at 5 GHz.
The total radio spectrum is relatively steep ($\alpha \simeq 0.9$;
$S(\nu) \propto \nu^{-\alpha}$) between 1.4 GHz (Kuhr et al.1981) 
and 5 GHz (Gregory \& Condon 1991), but it flattens
at lower frequencies ($\alpha \simeq 0.65$) between 178 and
1400 MHz (Kuhr et al.1981). 
Such a spectral trend may indicate the presence of a radiative
break in the energy distribution of the relativistic
electrons. We find that the spectrum can be reproduced by 
a {\it continuous injection} model (e.g. Kardashev 1962).
We obtain a break frequency $\nu_{\rm b}$= 0.5--1 GHz and an injection
spectral index of the electrons $\delta \sim 2.1-2.3$
($f(\gamma)\propto \gamma^{-\delta}$). 
This scenario is consistent with the radio, optical and 
X--ray data of the northern 
hot spot of 3C 295 (Harris et al.2000) which are well accounted for
by the combined synchrotron and SSC processes under the
assumption of an electron injection index $\delta \sim 2.3$
(Brunetti 2001). 
Here it should be noticed that the adiabatic expansion, suffered by the
electrons moving from the hot spot region to the lobes, 
would shift the energy of the electrons 
required to emit the observed
optical emission of the hot spot via the SSC
process from $\gamma \sim 600$ to $\gamma$\ltsim 200 (i.e., expanding
by a factor \gtsim 3).
These resulting low energies are those required to upscatter
the quasar far--IR photons into the X--ray band.

{\it c) Radiation pattern and luminosity of the 
nuclear source}. 
According to the unification scheme (Barthel 1989;
Pier \& Krolik 1992) we assume that the far--IR/optical
photons are isotropically emitted within two opposite cones
of half opening angle $\theta_{\rm C}=40^o$ and with the 
axis coincident with
the radio axis.
Almost all the volume occupied by the radio lobes,
as defined by the radio contours (Fig.3), are contained
within these cones.
The corresponding isotropic luminosity of the hidden quasar is
$L_{\rm Q} \sim 10^{46}$erg s$^{-1}$ as estimated in Sect. 3.1 by
assuming the SED of Sanders et al.(1989).

The additional effect due to the far--IR/optical
beamed radiation emitted by the nucleus is not
considered in the calculation. However, as discussed
in Brunetti et al.(1999), the beamed radiation
intercepts only a small fraction of the
lobes ($\theta_{\rm beam} \sim 5^o$) so that, in general,
it would not contribute more than 10-20$\%$ to the
IC scattered X--ray luminosity.
The low photon statistics do not
allow a precise mapping of the X--ray distribution in the lobes. 
However, from the analysis reported in Sect. 2.2 we derive
a rough lower limit, $\theta > 30^o$, to the total
angular distribution (referred to the central source)
of the X--rays in the northern lobe by simply dividing
the lower limit to the extension perpendicular to the radio axis 
by the extension of the X--ray lobe along the radio axis.
This also provides direct observational evidence that beamed nuclear 
radiation ($\theta_{\rm beam} \sim 5-10^o$) does not represent an
important source of IC X--rays from 3C 295.

The only free parameter of the model is the inclination
$\theta_{\rm ax}$ of the radio axis on the plane of the sky 
(Fig.A1).

With the previous assumptions we obtain the following
results :

\begin{figure}
\resizebox{\hsize}{!}{
\includegraphics{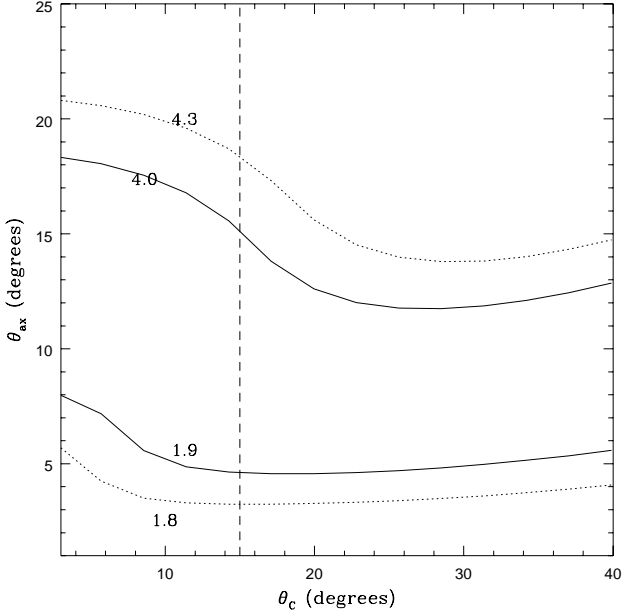}}
\caption[]
{The value of $\theta_{\rm ax}$ required to match the
observed lobe X--ray asymmetry 
is reported as a function of $\theta_{\rm C}$.
The lower and upper limits on the flux ratio between the X--ray lobes
are given in the panel at 90\% (dotted line)
and 68\% (solid line) confidence level.
The dashed line represents the lower limit on $\theta_{\rm C}$ as derived by
the observed distribution of the X--rays in the
northern lobe.}
\end{figure}

{\it 1) Reproducing the X--ray asymmetry of the lobes}.
If the radio galaxy lies on the plane of the sky, 
the implied X--ray ratio between the two radio 
lobes would be 1.3, mostly due to the
fact that the northern lobe is closer to the emitting
nucleus.  
In order to reproduce the observed interval
for the ratio of X--ray lobes, a range of $\theta_{\rm ax}$
between 6 and 13$^o$ (68\% confidence level)
is required by the model.
The northern lobe would be further away from us
so that its X--ray brightness is enhanced by
predominant back scatterings.
This moderate inclination is consistent with 
the absence of a clear radio jet in the radio
images of 3C 295 published in the literature
(Taylor \& Perley 1991,92; Akujor et al.1994).
However, a faint radio jet has been recently 
discovered in the southern lobe
by a deep MERLIN observation
(P.Leahy, private communication) thus 
providing independent confirmation that this
lobe is the near one as required by the model.

As shown in Fig.6 this result does not depend
on the assumed value of $\theta_{\rm C}$ since for 
$\theta_{\rm C}>15^o$ the allowed
interval of $\theta_{\rm ax}$ is well established
and stable in the range 5--15$^o$
(68\% confidence level).

Furthermore, we find that almost independently from 
the assumptions on the nuclear emitting pattern, 
the expected trend in X--ray brightness
between 0--0.5 and 1--1.5 arcsec from the nucleus
in the northern lobe  
is rather flat, which is again
consistent with the observations.

\begin{figure}
\resizebox{\hsize}{!}{
\includegraphics{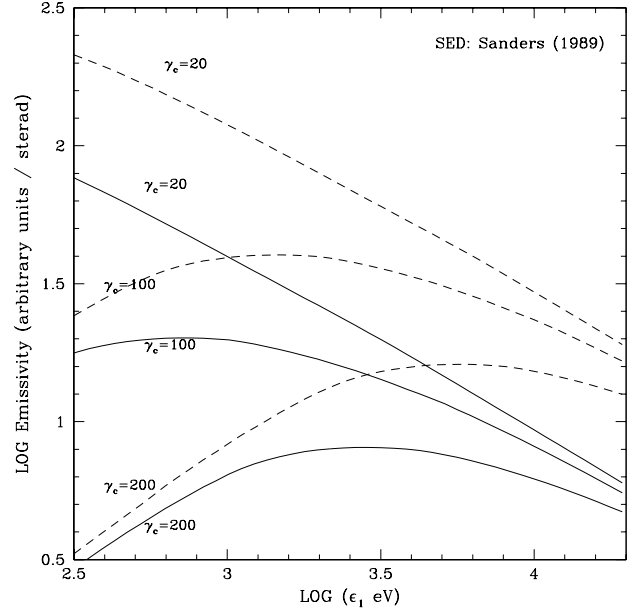}}
\caption[]
{The rest frame emissivity from the IC scattering of
nuclear photons by relativistic electrons is reported
for different low energy cut--offs as in the panel.
We have assumed two scattering angles ($\theta_{\rm SC}$):
$\mu=\cos \theta_{\rm SC}=0$ (solid lines) and $\mu=-1$ 
(dashed lines); $\delta=2.3$ and
the quasar SED of Sanders (1989) have been assumed.}
\end{figure}

{\it 2) Constraining the low energy spectrum of the 
relativistic electrons}.
In addition to the information 
on the spectrum of the relativistic 
electrons at higher energies 
as obtained by radio observations, the spectrum
of $\gamma \sim 100-300$ electrons can be constrained
by comparing the model predictions with the {\it Chandra} data.
In particular, the presence of a possible low energy cut--off
($\gamma_{\rm c}$) can be tested.
As an illustrative example, in Fig.7 
we report the IC spectra due to the anisotropic
IC scattering
of nuclear photons, by assuming the quasar SED given by
Sanders et al.(1989), as a function of the low
energy cut--off.
The curves are drawn for two values of the
scattering angle $\theta_{\rm SC}$ (see Fig.A1): 
$\mu = \cos \theta_{\rm SC}
= 0$ and $\mu =-1$ (back scattering case); 
since the typical 
electron energy involved in up--scattering 
the photon energy from $\epsilon_0$
to $\epsilon_1$ is 
$\gamma \sim \sqrt{\epsilon_1/(1-\mu)\epsilon_0}$,  
the back scattering case is more sensitive to the
presence of a low energy break in the electron 
spectrum.
We remark that the IC spectra 
are sensitive to the shape of the assumed SED, so that 
the presence of a low energy cut--off in the electron
energy distribution is less evident in the case of 
a redder SED such as, for instance, 
that of 3C 48 (reported in Haas et al.1998).

\begin{figure}
\resizebox{\hsize}{!}{
\includegraphics{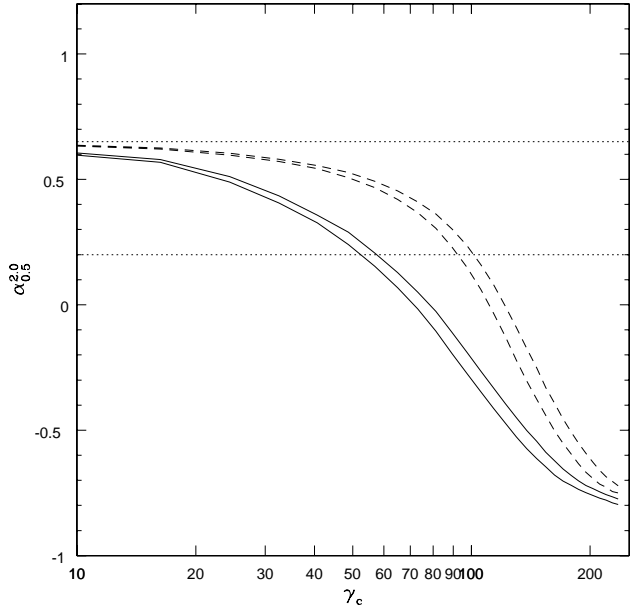}}
\caption[]
{The 0.5--2 keV spectral index 
expected in the case of the IC scattering of nuclear photons
is reported in the case of 3C 295.
In the calculation we have assumed both Sanders (1989)
(solid lines) and 3C 48 SEDs (dashed lines).
The plots are given for $\delta=2.3$, $\gamma_{\rm b} > 1000$  
and are K-corrected for $z=0.46$.
For each SED we have reported two different curves
representing the minimum and maximum values of the
spectral index as expected by the model; the calculations
are performed with the model parameters 
within the following ranges:
$\theta_{\rm C}=15-40^o$, $\theta_{\rm ax}=5-15^o$ and
$\delta=2.1-2.3$.
The limits on the observed 0.5--2 keV spectral
index from our {\it Chandra} data 
analysis are also reported (dotted lines).}
\end{figure}

In Fig.8 we report the model 0.5--2 keV spectral index as a function
of the low energy cut--off in the electron spectrum by
assuming both Sanders (1989) and 3C 48 SEDs.
By comparing the theoretical results with the limits given by
the {\it Chandra} data we obtain
an upper limit to $\gamma_{\rm c} <  100$.
For the sake of completeness in the same figure we
also show that this result is essentially independent,
as it should be, from the precise values assigned to
$\theta_{\rm C}$, $\theta_{\rm ax}$ and $\delta$.
Likewise, the asymmetry discussed in point 1) above does not
significantly depend on $\gamma_{\rm c}$.

{\it 3) Constraining the energetics of the radio lobes.}
The IC scattering of the nuclear photons can be further
used to test the 
the minimum energy condition in the radio lobes once the quasar
luminosity is fixed. 
By matching the expected IC flux from the radio lobes
with the {\it Chandra} 0.1--10 keV
flux from the lobes after the subtraction of the 
hot spot contribution ($\Gamma=1.6$ in Sect. 2.2)
we obtain:

\begin{equation}
{{ B_{\rm IC} }\over {B_{\rm eq}(\gamma_{\rm c})}} \sim
\left[ {{0.8 \eta(\gamma_{\rm c}) L_{\rm Q,46}}\over
{(1+k)^{ {{\delta+1}\over{\delta+5}} } }} \right]^{ {2\over{\delta+1}} }
\left( {{ \gamma_{\rm c}}\over{
50}} \right)^{ {{2(\delta-2)}\over{\delta+5}} }
\label{bic}
\end{equation}

\noindent
where $B_{\rm IC}$ is the magnetic field intensity necessary to account for
the X-ray flux, $B_{eq}$ ($\simeq 120 \mu$G) 
is the equipartition field derived by 
taking into account the presence of low 
energy relativistic particles ($\gamma \geq \gamma_{\rm c}$) 
not revealed by the synchrotron radio emission
(e.g. Brunetti et al.1997), $\eta(\gamma_{\rm c})$ 
is the ratio between the IC luminosity as a function of 
$\gamma_c$ and that with
no low energy cut-off in the electron spectrum and $k$
is the ratio between the energy densities of the protons 
and of the electrons.

\noindent
Eq.(\ref{bic}) is reported in Fig.9 for the representative
values $\delta=2.3$ and $k=1$, and for two 
values of the hidden quasar luminosity.
With $\gamma_c < 100$ and 
$L_{\rm Q} \sim 1 \cdot 10^{46}$erg s$^{-1}$ one finds
that the magnetic field estimated by the IC model is
within a factor $\sim 2$ from the value calculated
under minimum energy assumption.

\begin{figure}
\resizebox{\hsize}{!}{
\includegraphics{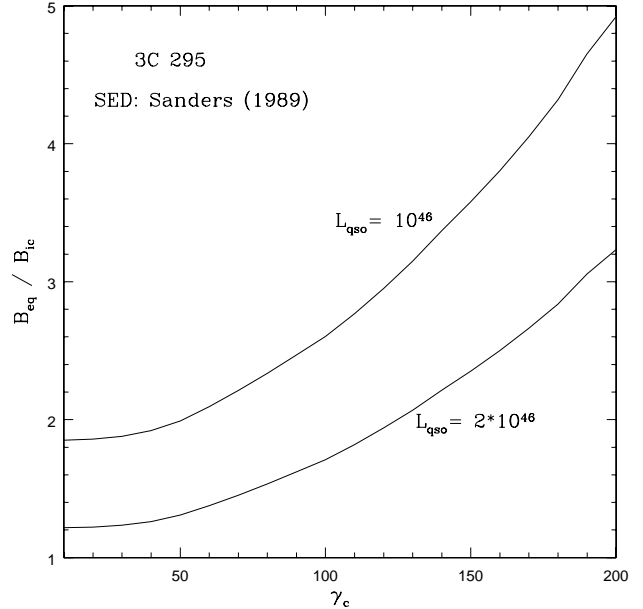}}
\caption[]
{The ratio between the equipartition magnetic
field intensity ($B_{\rm eq}$) and that estimated 
from the IC scattering is reported as a function 
of the low energy
cut--off $\gamma_{\rm c}$.
The calculation are performed for two bolometric
far--IR/optical luminosities of the hidden quasar
as shown in the panel.
The adopted SED is that from Sanders (1989).}
\end{figure}
 
We wish to stress that while the results discussed in 1) and 2) above
are essentially independent from the precise value of 
$\theta_{\rm C}$, this
obviously does not apply to the conclusions reached in 3). 
A smaller
value of $\theta_{\rm C}$ implies that the quasar radiation cone 
intercepts only 
a fraction of the radio lobe volume and to match the observed extended
X-ray flux one would have to assume either a more powerful quasar or a
larger number of relativistic electrons 
(bringing the radio lobes very much out
of equipartition) or both. 
The size of this effect is shown in Fig.10.
One derives that with $\theta_{\rm ax} \sim 10^o$ 
a change of $\theta_{\rm C}$ from 40$^o$ (adopted
in our model) to 30$^o$ does not significantly affect our conclusions;
instead, the adoption of a $\theta_{\rm C}$ value close to the lower 
limit of
15$^o$ would increase the energy requirements 
(hidden quasar luminosity and/or electron number density)
by one order of magnitude.
For completeness in Fig.10 we have also drawn the curve for
$\theta_{\rm C}$ = 5$^o$, 
representative of the beamed radiation cone of blazar
type objects, showing that the energy requirements are increased by
about two orders of magnitude. Although
not relevant for the present
discussion about 3C 295, the expected contribution of beamed
radiation to the X-ray brightness via the anisotropic IC scattering
may be detected by Chandra observations of well resolved powerful
radio galaxies.

\begin{figure}
\resizebox{\hsize}{!}{
\includegraphics{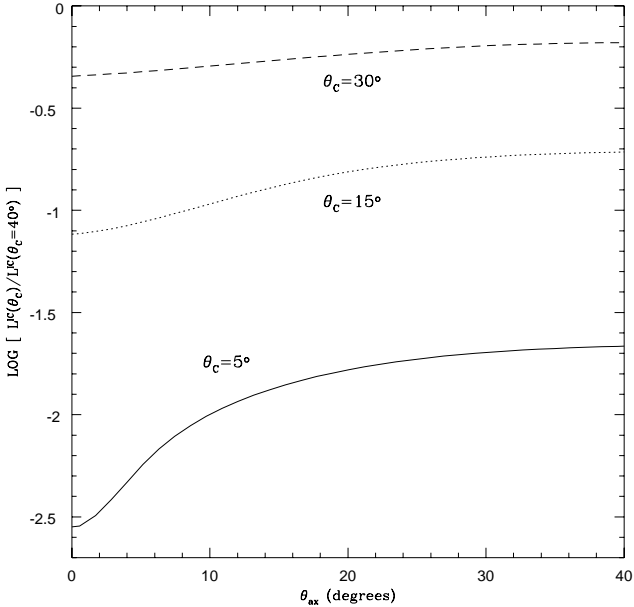}}
\caption[]
{The ratio between the IC luminosity of the northern
lobe of 3C 295 calculated with
different $\theta_{\rm C}$ (as given in the panel)
and that with $\theta_{\rm C}=40^o$ is
reported as a function of $\theta_{\rm ax}$.}
\end{figure}

\section{Discussion and conclusions}

In this paper we have re--analyzed the {\it Chandra}
observation of the powerful radio galaxy 3C 295
in order to study the extended X--ray
emission associated with the radio lobes.

We show that the 0.1--2 keV image is not affected by the strongly
absorbed ($N_{\rm H} \sim 10^{23}$cm$^{-2}$) nuclear source and
presents extended emission with a double lobed structure,
the northern lobe being a factor 2--4 stronger than the
southern one, strongly correlated with the morphology of 
the radio lobes, so that a non--thermal origin of the
X--rays is by far the most favoured.

We have examined different non--thermal scenarios  
including SSC, IC scattering of CMB photons and 
of photons emitted by a powerful hidden quasar.
The energy density due to the nuclear photons
largely dominates (10--200 times) 
that due to CMB and synchrotron
photons in the regions where extended X--ray emission
is detected, while the energy density due to
synchrotron photons outweighs the other contributions
in the northern hot spot region where, indeed, a powerful
compact X--ray source is detected.

By assuming an average constant ratio
between the relativistic electron and magnetic field energy 
densities in the
radio lobes, we show that the X--ray lobe 
asymmetry and surface brightness can only be
accounted for by the IC scattering of
the nuclear photons confirming the general 
model originally proposed by Brunetti et al.(1997).
A recent detection of a weak radio jet pointing
toward the southern lobe (Leahy, private com.) provides
additional observational evidence that the northern lobe 
is the farthest as required 
by our model.

Finally, we have applied the model to constrain the electron spectrum
and energetics in the lobes of 3C 295. 
The main results are as follows:

-- The X--ray flux ratio between the northern and southern lobes can
be accounted for by supposing that the main radio axis of the source
is slightly inclined ($\sim 5-13$ deg) with respect to the 
plane of the sky.
This result is fairly independent on the assumptions on the
nuclear emitting pattern.
  
-- By adopting the power law energy distribution of the 
relativistic electrons derived from the low frequency synchrotron
radio emission, we can set an upper limit on the possible presence 
of a low energy cut--off $\gamma_{\rm c}$. 
This is because the soft X--rays 
are mainly produced by the up--scattering of the nuclear IR photons
with electrons of energy $\gamma \sim 100-300$, much lower than that 
producing the radio emission by the synchrotron process 
($\gamma$\gtsim$10^4$), so that the computed spectra are sensitive to the
location of $\gamma_{\rm c}$. 
By comparing the soft X-ray spectra (0.15-3 keV in the source frame)
obtained from the anisotropic IC scattering with the
observational bounds imposed by the {\it Chandra} observation we find
$\gamma_{\rm c} < 100$, reaching the interesting conclusion that the power 
law electron spectrum defined by the radio emission can be extrapolated 
downward in energy by at least two orders of magnitude.

-- The average magnetic field intensity $B_{\rm IC}$ required to fully 
account for the lobe X-ray emission depends on the hidden quasar 
far--IR/optical luminosity $L_{\rm Q}$,
on the location of the low energy cut-off $\gamma_c$ and the ratio 
($k$) between relativistic protons and electrons. 
With $\gamma_{\rm c} < 100$ and 
$k = 1$, we find that $B_{\rm IC}$ is a factor 1.2--3.5 smaller than the 
equipartition value (obtained by taking into account the low energy 
particles) for $L_{\rm Q} = 0.5 - 2 \times 10^{46}$ erg s$^{-1}$
(note that the value 3.5 is calculated for $\gamma_{\rm c}=100$).
Therefore it may very well be that in this 
source $B_{\rm IC} \simeq B_{\rm eq}$ (implying a reasonable value
$L_{\rm Q} \simeq 2.5 \times 10^{46}$ erg s$^{-1}$)
\footnote{
After this paper was revised, ISOPHOT measurements of
dust emission from 3C 295 have been published by
Meisenheimer et al.(2001).
The 5--100$\mu m$ isotropic luminosity derived from the published 
fluxes is $\sim 2.5 \cdot 10^{46}$erg s$^{-1}$ ($H_0=50$,
$q_0=0.5$) pointing to the higher values of the quasar
bolometric luminosity adopted in our model
and to the existence of equipartition condition 
in the lobes.}.

Here, for comparison, it should be stressed that
in order to match the diffuse X--ray flux,  
either the SSC process and the IC scattering of 
CMB photons would require an electron number
density (and energetics) about 200 
times larger than the equipartition value.

Moderate deviations from equipartition  
within a factor of $\sim 3$ in magnetic field intensity 
($B_{\rm IC} < B_{\rm eq}$) 
have been suggested in the case of radio galaxies 
much more extended than 3C 295 (e.g. Cen B, Tashiro et
al.1998; 3C 219, Brunetti et al. 1999; Fornax A, Tashiro et al.2000) 
so that one might speculate that a moderate 
particle dominance could be linked to the 
time evolution of the radio galaxies, a condition recently invoked by 
Blundell \& Rawlings (2000) to match dynamical and
radiative ages in radio sources.
Of course, this scenario cannot be tested by present X--ray studies 
but only by the future {\it Chandra} and 
XMM--{\it Newton} observations.

\begin{acknowledgements}

We thank G.B. Taylor and R.A. Perley for providing
the VLA radio image of 3C 295 in Fig.3 and A. Comastri
for useful discussions.
We warmly thank J.P. Leahy who communicate to
us his discovery of a faint
radio jet in the southern lobe of 3C 295 
during the preparation of this paper.
We also acknowledge the anonymous referee whose comments
have improved the presentation of the paper.
This work was partly supported by the Italian Ministry for
University and Research (MURST) under grant Cofin98-02-32, 
and by the Italian Space Agency (ASI).

\end{acknowledgements}

\appendix{}
\section{Spatial asymmetries from anisotropic IC
scattering}

\begin{figure}
\includegraphics{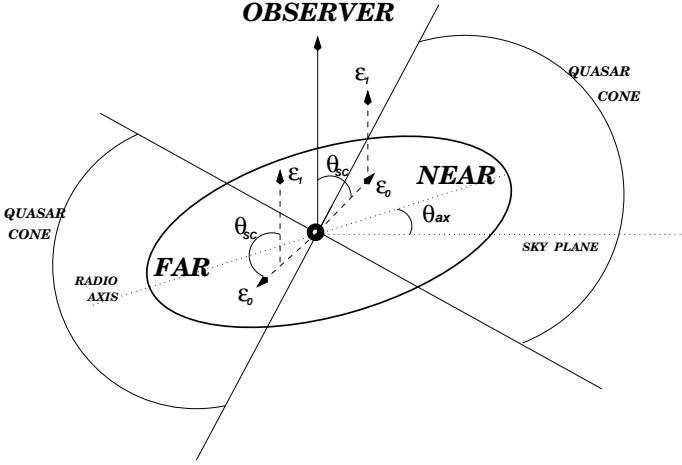}
\vspace{7 cm}
\caption{A schematic representation of the 
anisotropic IC scattering occurring 
in a radio galaxy approximated with a prolate
ellipsoid (thick solid line) is shown.
The nuclear photons ($\epsilon_0$) and scattered ($\epsilon_1$)
photons are indicated with the dashed arrows, while
radio axis and sky plane are represented with dotted
lines.}
\end{figure}

By making use of semi--analytic anisotropic inverse
Compton formulae, in this Appendix we give a general formalism
to calculate the ratio between the X-ray intensities from
the two lobes.

\noindent
The general anisotropic IC emissivity 
per unit solid angle of incoming photons 
($\epsilon_0$) scattered into an angle 
$\cos \theta_{\rm SC} = \mu$,
is derived in Brunetti (2000).
Here we consider the ultra--relativistic limit
(i.e. $x \equiv \epsilon_1/\epsilon_0 >> 1$) 
of the equations
derived for an isotropic electron momentum distribution, 
$N_e(\gamma,\Omega)= K_e f(\gamma)/ 4\pi$, 
in the Thompson case, i.e. :

\begin{equation}
j(\mu, x) = C_0 K_{\rm e} N^{\rm ph}(\epsilon) x^2
\left\{ {{2{\cal I}_2(x)}
\over{x}}
-{{2 {\cal I}_4(x) }\over{ 1- \mu }} +
{{ x {\cal I}_6(x) }\over{ (1- \mu )^2}}
\right\}
\label{j1}
\end{equation}

\noindent
where $C_0$ is a constant, 
$N^{\rm ph}(\epsilon)$ is the 
incoming photon number density, 

\begin{equation}
{\cal I}_s (x) = \int_{\gamma_{\rm min}}
f_{ {\bf r} }(\gamma) \gamma^{-s}
\label{is}
\end{equation}

\noindent
and $\gamma_{\rm min}= \{ x/[2 (1-\mu) ] \}^{1/2}$.
In the case of 
a simple power law energy distribution  
$f(\gamma) = \gamma^{-\delta}$ 
Eq.(\ref{j1})
approaches $j \propto  (1-\mu)^{(\delta + 1)/2}$
and the ratio between the emissivities  
at different $\mu$ does not depend on $x$. 
However, in general the energy distribution 
of the relativistic electrons undergoing 
re--acceleration processes, radiative, 
adiabatic and Coulomb losses differs from
a simple power law so that such a ratio can 
depend on the observing frequency
and on $N^{\rm ph}(x)$. 

The IC luminosity per unit solid angle
from the volume $V$ is :

\begin{equation}
L^{\rm IC} = \int d\epsilon \int_{V} d\mu d\phi r^2 dr
j(\mu, x, {\bf r}) 
\label{l1}
\end{equation}

In our model the radio lobes are illuminated by
the nuclear source, so that 
$N^{\rm ph}(\epsilon,{\bf r})=
N^{\rm ph}(\epsilon,\mu,\phi)/4\pi c r^2$; 
$N^{\rm ph}(\epsilon,\mu,\phi)$ takes
into account the pattern of the emitted 
nuclear photons.
From Eqs.(\ref{j1}) and (\ref{l1}) the luminosity
from a given radio volume is:

\begin{eqnarray}
L^{\rm IC} = C_1 \int_{\epsilon} d\epsilon\, x^2
\int d\mu \int d\phi\,
N^{\rm ph}(\epsilon,\mu,\phi) \times  
\nonumber\\
\int dr \left\{ {{2{\cal I}_2(x,{\bf r})}
\over{x}}
-{{2 {\cal I}_4(x,{\bf r}) }\over{ 1- \mu }} +
{{ x {\cal I}_6(x, {\bf r}) }\over{ (1- \mu )^2}}
\right\} 
\label{l2}
\end{eqnarray}

\noindent
with $C_1$ a constant.

\begin{figure}
\resizebox{\hsize}{!}{\includegraphics{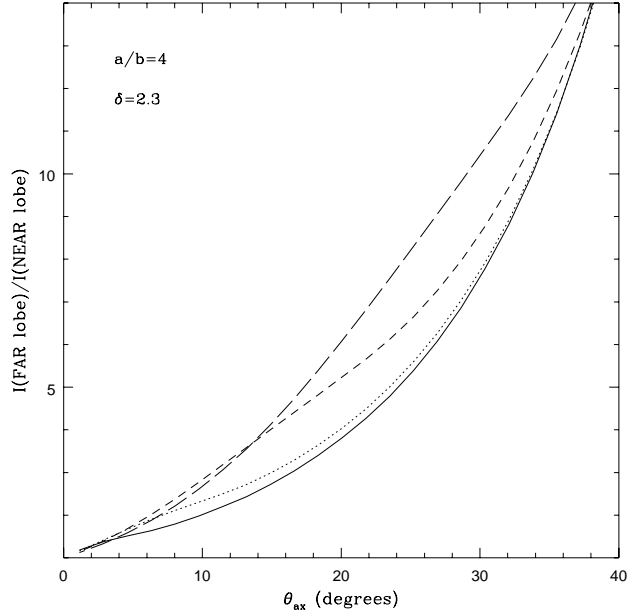}}
\caption[]
{The ratio between far and near X--ray lobe
at $\epsilon_1= 1$ keV is reported as a function of
the inclination angle $\theta_{ax}$ (Fig.A2).
The results are reported for different half opening
angles of the quasar emitting cone (Eq.\ref{boundary}):
$\theta_c$= 0.1 (solid line), 0.2 (dotted line),
0.4 (short dashed line), 0.6 rad (long dashed line).
We have assumed a monochromatic incoming photon beam
of energy $\epsilon = 0.02$ eV, a radio volume given
by an ellipsoid with ratio between the axis $a/b =4$, 
$\delta=2.3$, $\gamma_{\rm c}=1$, and $\gamma_{\rm b}= 10^4$.}
\end{figure}

\begin{figure}
\resizebox{\hsize}{!}{\includegraphics{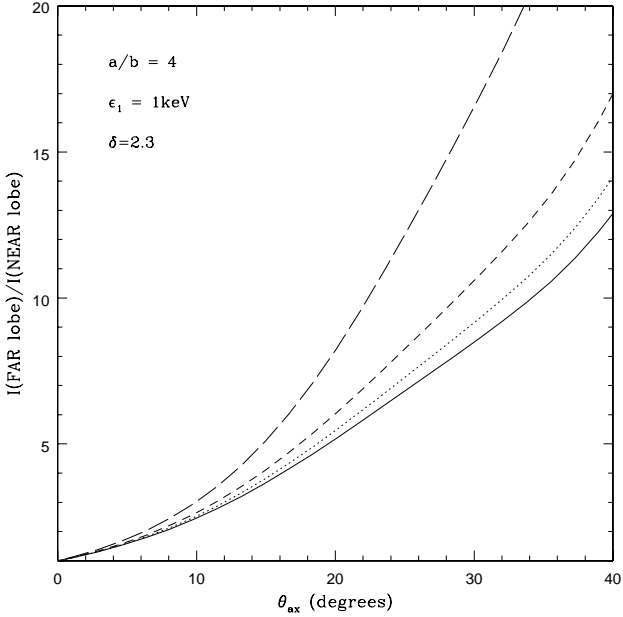}}
\caption[]
{The ratio between far and near X--ray lobe
at $\epsilon_1= 1$ keV is reported as a function of
the inclination angle $\theta_{\rm ax}$ (Fig.A3).
The results are reported for different break
energies: $\gamma_{\rm b}$= $\infty$ (solid line), 
1000 (short dotted line), 500 (short dashed line), and 
300 (long dashed line).
We assume $\theta_{\rm c}$=0.6 rad and the other parameters as in Fig.A2.}
\end{figure}

\noindent
Once the electron energy distribution $f(\gamma,{\bf r})$
is given, Eq.(\ref{l2}) yields a general formalism to calculate
anisotropic IC emissions per unit solid angle and at
a given frequency.

To simplify the scenario in the case of the radio sources, 
we approximate the radio volume with a prolate
ellipsoid of major and minor semiaxis $a$ and $b$
respectively (Fig.A1) and consider 
an uniform electron energy distribution.
From Eq.(\ref{l2}) one obtains the IC luminosity from
one lobe:

\begin{eqnarray}
L^{\rm IC} = C_2 \int d\epsilon\, x^2
\int d\mu 
\left\{ {{2{\cal I}_2(x)}
\over{x}}
-{{2 {\cal I}_4(x) }\over{ 1- \mu }} +
{{ x {\cal I}_6(x) }\over{ (1- \mu )^2}}
\right\} \nonumber\\
\int_{e_1}^{e_2}
{{ N^{\rm ph}(\epsilon,\mu,\phi) d\phi}\over{
\sqrt{ \mu^2 \left( 
\sin^2 \phi \, k_1(\mu) + \sin \phi \, k_2(\mu)
+k_3(\mu) \right) } }}
\label{l3}
\end{eqnarray}

\noindent
where the integral in $\mu$ is 
performed in the interval $[-1,1]$, while  
the extremes 
$e_1$ and $e_2$ are 0 or $\pi$ and $\pi$ or $2 \pi$ 
in the case of the near and far lobe 
respectively (Fig.A1); $C_2$ is a constant.
The functions $k_i$ in Eq.(\ref{l3}) are:

\begin{equation}
k_1(\mu)=
{{ 1-\mu^2}\over{\mu^2}} 
\tan^{-2}(\theta_{\rm ax}) \left[ \left({b\over a} \right)^2
-1 \right]
\end{equation}

\begin{equation}
k_2(\mu)=
2 {{\sqrt{1-\mu^2}}\over{\mu}} \tan^{-1} (\theta_{\rm ax}) 
\left[ 1 - \left( {b \over a} \right)^2
\right]
\end{equation}

\begin{equation}
k_3(\mu)=
\tan^{-2}(\theta_{\rm ax}) + \left( {b \over a} \right)^2
+\left[ 1 + \tan^{-2}(\theta_{\rm ax}) \right] 
{{ 1-\mu^2}\over{\mu^2}} 
\end{equation}

\noindent
In the simple case of an uniformly emitting cone
pattern of half opening angle $\theta_{\rm c}$, one
has:

\begin{eqnarray}
N^{\rm ph}(\epsilon,\mu,\phi) =
\cases{
N^{\rm ph}(\epsilon) \,\,\, 
for \,\,\, (\mu,\phi) \in {\cal B}(\theta_{\rm c}\, , \, 
\theta_{\rm ax})  \cr
                                                    \cr
0 \,\,\,\,\, outside                                \cr
                                                    \cr}
\label{cone}
\end{eqnarray}

\noindent
where the boundary is defined as:

\begin{eqnarray}
{\cal B}(\theta_{\rm c}\, , \, \theta_{\rm ax}) \equiv 
\cases{
NEAR \,\, lobe \, \Rightarrow          \cr
{{\pi}\over 2}-\theta_{\rm ax} -\theta_{\rm c}
<\theta<
{{\pi}\over 2}-\theta_{\rm ax}+\theta_{\rm c} \, ,         \cr
{{\pi}\over 2}-\theta_{\rm c}
<\phi<
{{\pi}\over 2}+\theta_{\rm c}                \cr
FAR \,\, lobe \, \Rightarrow           \cr
{{\pi}\over 2}+\theta_{\rm ax} -\theta_{\rm c}
<\theta<
{{\pi}\over 2}+\theta_{\rm ax}+\theta_{\rm c} \, ,         \cr
{{3 \pi}\over 2}-\theta_{\rm c}
<\phi<
{{3 \pi}\over 2}+\theta_{\rm c}              \cr}
\label{boundary}
\end{eqnarray}

\begin{figure}
\resizebox{\hsize}{!}{\includegraphics{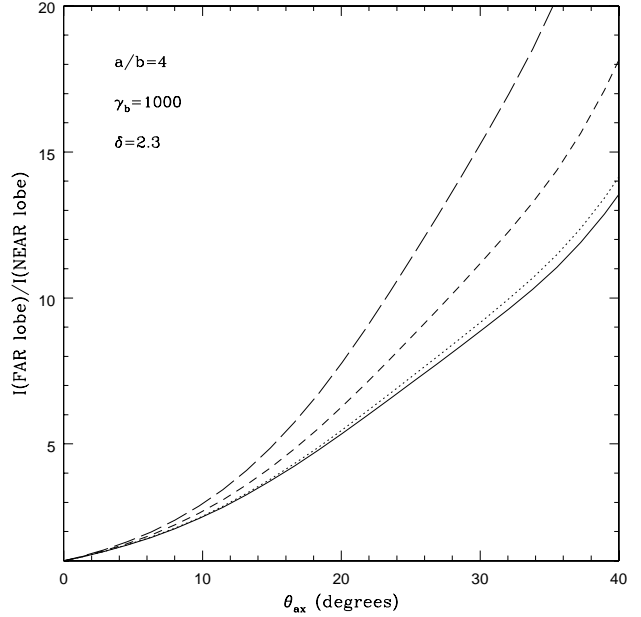}}
\caption[]
{The ratio between far and near X--ray lobe
for $\gamma_{\rm b}=1000$ is reported as a function of
the inclination angle $\theta_{\rm ax}$ (Fig.A2).
The results are reported for different 
scattering energies:
$\epsilon_1$= 200 eV (solid line), 
1000 eV (dotted line), 5000 eV (short dashed line),
10000 eV (long dashed line).
The other parameters are the same as in Fig.A3.}
\end{figure}

\noindent
In Fig.A2 we report the ratio between far and near lobe
X--ray luminosities as a function of $\theta_{\rm ax}$ and for
different opening angles ($\theta_{\rm c}$) of the quasar emitting cone : 
larger $\theta_{\rm c}$ give in general larger X--ray ratios 
($\theta_{\rm ax} < 40$ deg).
As a consequence, in the case of radio galaxies,
the IC scattering of the wide--angle nuclear radiation
emitted by the dusty torus, disk and broad line region, 
produces a larger brightness asymmetry than the IC scattering
of the beamed nuclear radiation.  

The effects on such ratio due to the presence of a break energy 
in the electron energy distribution is reported
in Figs.A3 and A4.

In the case $\gamma_{\rm c} << 100 << \gamma_{\rm b}$, one has 
$f(\gamma)=\gamma^{-\delta}$ so that the ratio between two emitting
volumes is independent on $x$ and the integrals in 
Eqs.(\ref{l3}) simplify. For each lobe one has:

\begin{equation}
L^{\rm IC}= C_3
\int \int 
{{ d\mu (1-\mu)^{ {{\delta+1}\over 2} } d\phi}\over{
\sqrt{ \mu^2 \left( 
\sin^2 \phi \, k_1(\mu) + \sin \phi \, k_2(\mu)
+k_3(\mu) \right) } }}
\label{l4}
\end{equation}

\noindent
the extremes of the integrals are
given by Eq.(\ref{boundary}); $C_3$ a constant.


\begin{thebibliography}{}

\bibitem[]{} Allen S.W., et al. 2000, MNRAS, submitted
\bibitem[]{} Akujor C.E., Luedke E., Browne I.W.A., et al., 
1994, A\&AS 105, 247
\bibitem[]{} Barthel P. D., 1989, ApJ 336, 606
\bibitem[]{} Bell A.R., 1978, MNRAS 182, 147
\bibitem[]{} Blundell K.M., Rawlings S., 2000, AJ 119, 1111
\bibitem[]{} Brinkmann W., Yuan W., Siebert J., 1997, 
A\&A 319, 413
\bibitem[]{} Brunetti G., 2000, APh 13, 105
\bibitem[]{} Brunetti G., 2001, in "Particles and Fields in Radio Galaxies", 
R.A.Laing and K.M.Blundell (Eds.), ASP Conf. Series, in press 
\bibitem[]{} Brunetti G., Setti G., Comastri A., 1997, A\&A 325, 898
\bibitem[]{} Brunetti G., Comastri A., Setti G., Feretti L., 
1999, A\&A 342, 57
\bibitem[]{} Canosa C.M., Worrall D.M., Hardcastle M.J.,
Birkinshaw M., 1999, MNRAS 310, 30
\bibitem[]{} Capetti A., Trussoni E., Celotti A., Feretti L., 
Chiaberge M., 2000, MNRAS 318, 493
\bibitem[]{} Carilli C.L., Perley R.A., Dreher J.W.,
Leahy J.P., 1991, ApJ 383, 554
\bibitem[]{} Carilli C.L., Perley R.A., Harris D.E., 1994, MNRAS 270, 
173 
\bibitem[]{} Carilli C.L., Kurk J.D.,
van der Werf P.P., Perley R.A., Miley G.K.,
1999, AJ 118, 2581
\bibitem[]{} Cash W., 1979, ApJ 228, 939
\bibitem[]{} Chartas G., Worrall D.M., Birkinshaw M., et al.
2000, ApJ 542, 655
\bibitem[]{} Crawford C.S., Fabian A.C., 1995, MNRAS 273, 827
\bibitem[]{} Crawford C.S., Fabian A.C., 1996, MNRAS 282, 1483
\bibitem[]{} Eilek J.A., Hughes P., 1990, in
'Astrophysical Jets', ed. Hughes P., 
Cambridge Univ. Press, 428 
\bibitem[]{} Elvis M., Wilkes B.J., McDowell J.C., et al., 
ApJS 95, 1
\bibitem[]{} Fabian A.C., Sanders J.S., Ettori S., et al.
2000, MNRAS 318, L65
\bibitem[]{} Fabian A.C., Crawford C.S., Ettori S., Sanders J.S.,
2001, MNRAS in press; astro-ph/0101478. 
\bibitem[]{} Feigelson E.D., Laurent-Muehleisen S.A., Kollgaard R.I.,
Fomalont E.B., 1995, ApJ 449, L149
\bibitem[]{} Gould R.J., 1979, A\&A 76, 306
\bibitem[]{} Gregory P.C., Condon J.J., 1991, ApJS 75, 1011
\bibitem[]{} Haas M., Chini R., Meisenheimer K., et al., 
1998, ApJ 503, L109
\bibitem[]{} Hardcastle M.J., Worrall D.M., 1999, MNRAS 309, 969
\bibitem[]{} Harris D.E., Grindlay J.E., 1979, MNRAS 188, 25
\bibitem[]{} Harris D.E., Nulsen P.E.J., Ponman T.J., et al., 
2000, ApJ 530, L81
\bibitem[]{} Henry J.P., Henriksen M.J., 1986, ApJ 301, 689 
\bibitem[]{} Kaiser C.R., Alexander P., 1997, MNRAS 286, 215
\bibitem[]{} Kaiser C.R., Alexander P., 1999, MNRAS 305, 707
\bibitem[]{} Kaneda H., Tashiro M., Ikebe Y., 
et al., 1995, ApJ 453, L13
\bibitem[]{} Kuhr H., Witzel A., Pauliny-Toth I.I.K., Nauber U., 
1981, ApJS 45, 367
\bibitem[]{} McNamara B.R., Wise M., Nulsen P.E.J., et al. 2000, 
ApJ 534, L135
\bibitem[]{} Meisenheimer K., Roser H.-J., Hiltner P.R., et al., 1989, 
A\&A 219, 63
\bibitem[]{} Meisenheimer K., Haas M., M\"{u}ller S.A.H., 
Chini R., Klaas U., Lemke D., 2001, A\&A in press;
astro-ph/0102333
\bibitem[]{} Mushotzky R.F., Scharf C.A., 1997, ApJ 428, L13
\bibitem[]{} Neumann D.M., 1999, ApJ 520, 87
\bibitem[]{} Pier E.A., Krolik J.H., 1992, ApJ 401, 99
\bibitem[]{} R\"{o}ser H.-J., Meisenheimer K., 1986, A\&A 154, 15
\bibitem[]{} Sanders D.B., Phinney E.S., 
Neugebauer G., Soifer B.T., Matthews K., 1989, ApJ 347, 29
\bibitem[]{} Sambruna R.M., Eracleous M., Mushotzky R.F., 1999,
ApJ 526, 60
\bibitem[]{} Schwartz D.A., Marshall H.L., Lovell J.E., et al.
2000, ApJ 540, 69L
\bibitem[]{} Taylor G.B., Perley R.A., 1991, AJ 101, 1623
\bibitem[]{} Taylor G.B., Perley R.A., 1992, A\&A 262, 417
\bibitem[]{} Tashiro M., Kaneda H., Makishima N., et al., 1998, 
ApJ 499, 713
\bibitem[]{} Tashiro M., Makishima N., Iyomoto N., Isobe N., Kaneda H.,
2000, ApJ L, in press; astro-ph/0010503
\bibitem[]{} Tsakiris D., Leahy J.P., Strom R.G., Barber C.R., 
1996, IAUS 175, 256
\bibitem[]{} van Bemmel I.M., Barthel P.D., Yun M.S., 1998, 
A\&A 334, 799
\bibitem[]{} Ueno S., Koyama K., Nishida M.,
Yamauchi S., Ward M.J., 1994, ApJ 431, L1
\bibitem[]{} Weaver K.A., 1993, PhD Thesis
\bibitem[]{} Wilson A.S., Young A.J., Shopbell P.L., 2000
ApJ L in press; astro-ph/0009308
\bibitem[]{} Wilson A.S., Young A.J., Shopbell P.L., 2001,
ApJ 546 in press; astro-ph/0008467
\bibitem[]{} Worrall D.M., 1999, in `Life Cycles of Radio 
Galaxies' eds. J.Biretta et al., New Astronomy Reviews
in press; astro-ph/9911056
\end{thebibliography}
\end{document}